\documentclass[apjl]{emulateapj}
\usepackage{color}

\usepackage{graphicx}

\begin{document}

\title{Direct Imaging Discovery of a Young Brown Dwarf Companion to an A2V Star}

\color{black}
\shorttitle{Discovery Images and Spectroscopy of HIP 75056Ab}
\shortauthors{Wagner et al.}
\author{Kevin Wagner\altaffilmark{1,2,3}$^{\star}$, D\'aniel Apai\altaffilmark{1,2,4}, Markus Kasper\altaffilmark{5}, Melissa McClure\altaffilmark{6}, Massimo Robberto\altaffilmark{7},\\ \& Thayne Currie\altaffilmark{8,9}}

\altaffiltext{1}{Steward Observatory, University of Arizona}
\altaffiltext{2}{NASA NExSS \textit{Earths in Other Solar Systems} Team}
\altaffiltext{3}{NASA Hubble/Sagan Fellow}
\altaffiltext{4}{Lunar and Planetary Laboratory, University of Arizona}
\altaffiltext{5}{European Southern Observatory}
\altaffiltext{6}{University of Amsterdam}
\altaffiltext{7}{Space Telescope Science Institute}
\altaffiltext{8}{Subaru Telescope / National Astronomical Observatory of Japan}
\altaffiltext{9}{NASA Ames Research Center}

\altaffiltext{$\star$}{Correspondence to: kwagner@as.arizona.edu}


\begin{abstract}

We present the discovery and spectroscopy of HIP 75056Ab, a companion directly imaged at a very small separation of 0$\farcs$125 to an A2V star in the Scorpius-Centaurus OB2 association. Our observations utilized VLT/SPHERE between 2015$-$2019, enabling low-resolution spectroscopy (0.95$-$1.65 $\mu m$), dual-band imaging (2.1$-$2.25 $\mu m$), and relative astrometry over a four-year baseline. HIP 75056Ab is consistent with spectral types in the range of M6$-$L2 and $T_{\rm eff}\sim$ 2000$-$2600 K. A comparison of the companion's brightness to evolutionary tracks suggests a mass of $\sim$20$-$30 M$_{Jup}$. The astrometric measurements are consistent with an orbital semi-major axis of $\sim$15$-$45 au and an inclination close to face-on (i$\lesssim$35$^o$). In this range of mass and orbital separation, HIP 75056Ab is likely at the low-mass end of the distribution of companions formed via disk instability, although a formation of the companion via core accretion cannot be excluded. The orbital constraints are consistent with the modest eccentricity values predicted by disk instability, a scenario that can be confirmed by further astrometric monitoring. HIP 75056Ab may be utilized as a low-mass atmospheric comparison to older, higher-mass brown dwarfs, and also to young giant planets. Finally, the detection of HIP 75056Ab at $0\farcs$125 represents a milestone in detecting low-mass companions at separations corresponding to the habitable zones of nearby Sun-like stars.

\end{abstract}

\section{Introduction}

Dozens of exoplanets and substellar companions have been directly imaged on orbits of $\sim$10$-$100 au around nearby young stars (e.g., \citealt{Bowler2016}; \citealt{Nielsen2019}; \citealt{Vigan2020}). These planets and brown dwarf companions are among the youngest known (e.g., \citealt{Macintosh2015, Meshkat2015, Keppler2018, Bohn2020}). Given their combination of mass and age, and also that their orbits and atmospheres can be readily characterized, directly imaged planets constitute important targets for studies of planet formation and planetary atmospheres. Future exoplanet imaging missions may also probe the habitable zones of Sun-like stars, as an Earth-analogue planet at 10 pc would appear at resolvable separations of $\sim$0$\farcs$1.

At masses of $\lesssim10~M_{Jup}$, most directly imaged planets represent the high-mass tail of planets formed via core accretion (e.g., \citealt{Kratter2010, Nielsen2019, Wagner2019}). Meanwhile, more massive objects ($\gtrsim$10-20~M$_{Jup}$) are likely formed predominantly via gravitational disk instability (e.g., \citealt{Boss1997, Forgan2018}). Low-mass objects (i.e., giant planets) formed by this process are likely rare, as the conditions required to trigger such instabilities require the presence of a significant amount of mass at the time of formation, which typically leads to the formation of a higher-mass brown dwarf or binary star \citep{Kratter2010, Forgan2018}. While some objects formed by either mechanism can occupy overlapping ranges of mass and semi-major axis distributions, these populations may display further differences, such as in their eccentricity distributions \citep{Bowler2020}, and in their atmospheric properties (e.g., \citealt{Spiegel2012}). These properties can now be measured with direct imaging, enabling probabilistic constraints on a companion's formation mechanism to be established (e.g., \citealt{Wagner2019}).


Directly imaged planets also provide some of the best available constraints on exoplanetary emission spectra and planetary atmospheres. Many directly imaged planets appear to be much redder than field objects of similar temperatures$-$indicating a significant cloud clover or a large amount of photospheric dust (e.g., \citealt{Currie2011, Madhu2011, Skemer2011, Bowler2017}). Disequilibrium chemistry also likely influences the observed features in the spectra of directly imaged planets (e.g., \citealt{Skemer2012}). Ultimately, a large body of observed exoplanet spectra$-$similar to the libraries of brown dwarf spectra with largely overlapping physical characteristics (e.g., \citealt{Manjavacas2019})$-$will enable a detailed understanding of their atmospheres as a function of mass, temperature, density, metallicity, age, and formation mechanism.

\begin{figure*}[t]
\epsscale{1.2}
\plotone{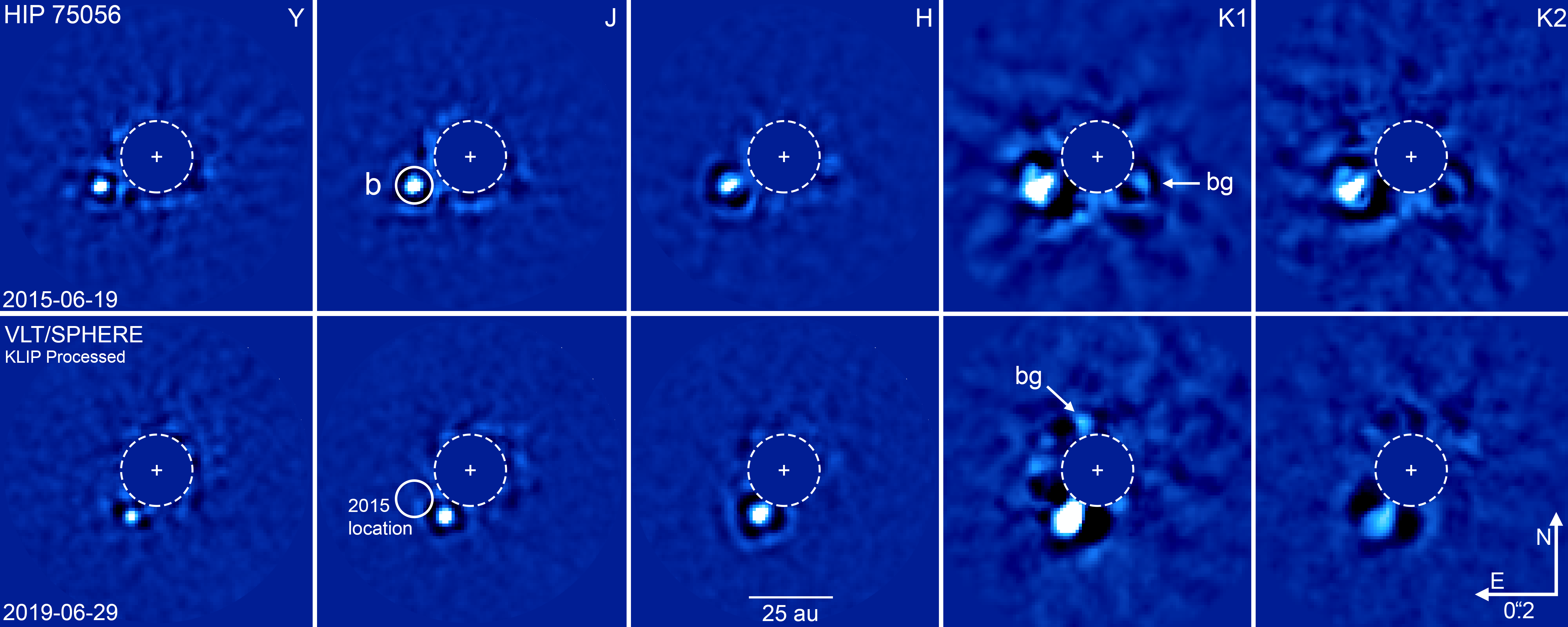}
\caption{SPHERE images of HIP 75056Ab from 2015 (top) and 2019 (bottom) processed with KLIP. The source is clearly detected at each end of the (1$-$2.25 $\mu$m) bandpass at a separation of $\sim$0$\farcs$155 (0$\farcs$125 in 2019). The object's motion with respect to HIP 75056A is consistent with orbital motion for a semi-major axis of $a\sim15-45$ au and is inconsistent with the expected motion of a background star. In the $K1$-band images, the relative positional change of the object labeled ``bg" illustrates the proper motion of HIP 75056.}
\end{figure*}

HIP 75056A is a $\sim$12 Myr old A2V star in the Upper-Centaurus-Lupus subgroup of the nearby Scorpius-Centaurus OB2 association (\citealt{deZeeuw1999, Pecaut2016, Gagne2018}). The primary star is orbited by a low-mass $\sim$0.3 M$_{\odot}$ star at an angular separation of 5$\farcs$2 \citep{Kouwenhoven2007}, or a projected separation of $\sim$650 au at the system's distance of 126$\pm$2 pc \citep{GDR2}. We observed HIP 75056A as the 25th target in our Scorpion Planet Survey (PI: Apai), which aims to establish the frequency of wide-orbit giant planets, and also to reveal new directly imaged companions to be utilized in future studies of giant planet formation and exoplanetary atmospheres (Wagner et al., \textit{in prep.}). Here, we report the discovery of a substellar companion around HIP 75056A at a very small angular separation of $0\farcs$125. We present an initial characterization of the companion's physical properties, and discuss the significance of its discovery.
\section{Observations and Data Reduction}

We observed HIP 75056A on 2015-06-19 and 2019-06-29 with the Spectro-Polarimetric High-contrast Exoplanet Research Experiment (SPHERE: \citealt{Beuzit2019}) on the Very Large Telescope (VLT). We utilized the \texttt{IRDIFS\_Ext} mode, which uses the Infrared Dual-band Imager and Spectrograph (IRDIS: \citealt{Vigan2012}) with the $K1K2$ filter combination (2.11 $\mu m$, 2.25 $\mu m$), simultaneously with the Integral Field Spectrograph (IFS: \citealt{Claudi2008}) in $Y-$ $H$-bands (0.95$-$1.65 $\mu m$). Our observations utilized the \texttt{N\_ALC\_YJH\_S} coronagraph, which enables observations at very small angular separations from the obscured star with $\sim$90\% transmission at $\sim$0$\farcs$125$-$0$\farcs$15 (SPHERE User Manual, v15).\footnote{\url{https://www.eso.org/sci/facilities/paranal/instruments/sphere}} On 2015-06-19, we obtained $\sim$21 minutes of observations (with detector integration times of 16 and 32 seconds for IRDIS and IFS, respectively) covering $\sim$14$^\circ$ of field rotation with average seeing of 1$\farcs$0. On 2019-06-29, we obtained $\sim$28 minutes of observations (with 50\% shorter detector integration times) and covering a larger $\sim$40$^\circ$ of field rotation with slightly poorer average seeing of 1$\farcs$2.

We reduced the data utilizing our previously developed SPHERE pipelines (see \citealt{Kasper2015, Apai2016, Wagner2018, Gibbs2019}), which we briefly describe here. We followed standard data reduction steps including dark subtraction, bad pixel correction from the mean of the surrounding pixels, flat-field division, distortion correction \citep{Maire2016}, and star-centering. For the 2015 dataset, we corrected the derotation angle for the time synchronization error between SPHERE and VLT's internal clocks following \cite{Maire2016}. These steps were followed by simulated point source injections in a copy of the dataset for subsequent sensitivity analyses (see \S3.4). 

We identified and removed bad frames via calculating the cross-correlation function of each image with respect to the median, and removed those with a cross-correlation value less than 0.85 for the IRDIS data and 0.95 for the IFS data. To remove remaining variations in the background we subtracted the mode of each column and row and then subtracted a 13 pixel median-smoothed version (9 pixels for the IFS images) of each image from itself. We modelled and subtracted the point spread function (PSF) of HIP 75056A with classical angular differential imaging (ADI: \citealt{Marois2006}) and projection onto eigenimages via Karhunen-Lo\`eve Image Processing (KLIP: \citealt{Soummer2012}), where we've specifically utilized the adaptation in \cite{Apai2016}. Finally, we combined the images using the noise weighting approach in \cite{Bottom2017}.

For the IRDIS data, we modeled the PSF with KLIP using the first two eigenvectors and eigen images in an annulus from 5$-$35 pixels (0$\farcs$06$-$0$\farcs$43) and no angular rotation criteria. For the IFS data, we first reduced the images with KLIP by generating a PSF basis from the images within the same wavelength (i.e, angular differential imaging mode, or ADI-KLIP) with two eigenvectors in four annular segments (of 90$^o$ azimuthal width and from 7$-$50 pixels, or 0$\farcs$05$-$0$\farcs$37), and with a minimum angular rotation criteria of 0.5 $\lambda$/D at 0$\farcs$18, or $\sim$8$^\circ$. We then processed the images a second time with KLIP in spectral differential imaging mode (SDI-KLIP), which utilizes the 39 spectral channels within an individual exposure as the PSF basis. We used four eigenvectors and images in the same region and with a minimum spectral magnification criteria of 1.5 $\lambda$/D at 0$\farcs$18, or about four spectral channels separation from the target channel.


\begin{figure*}[htpb]
\epsscale{1.15}
\plottwo{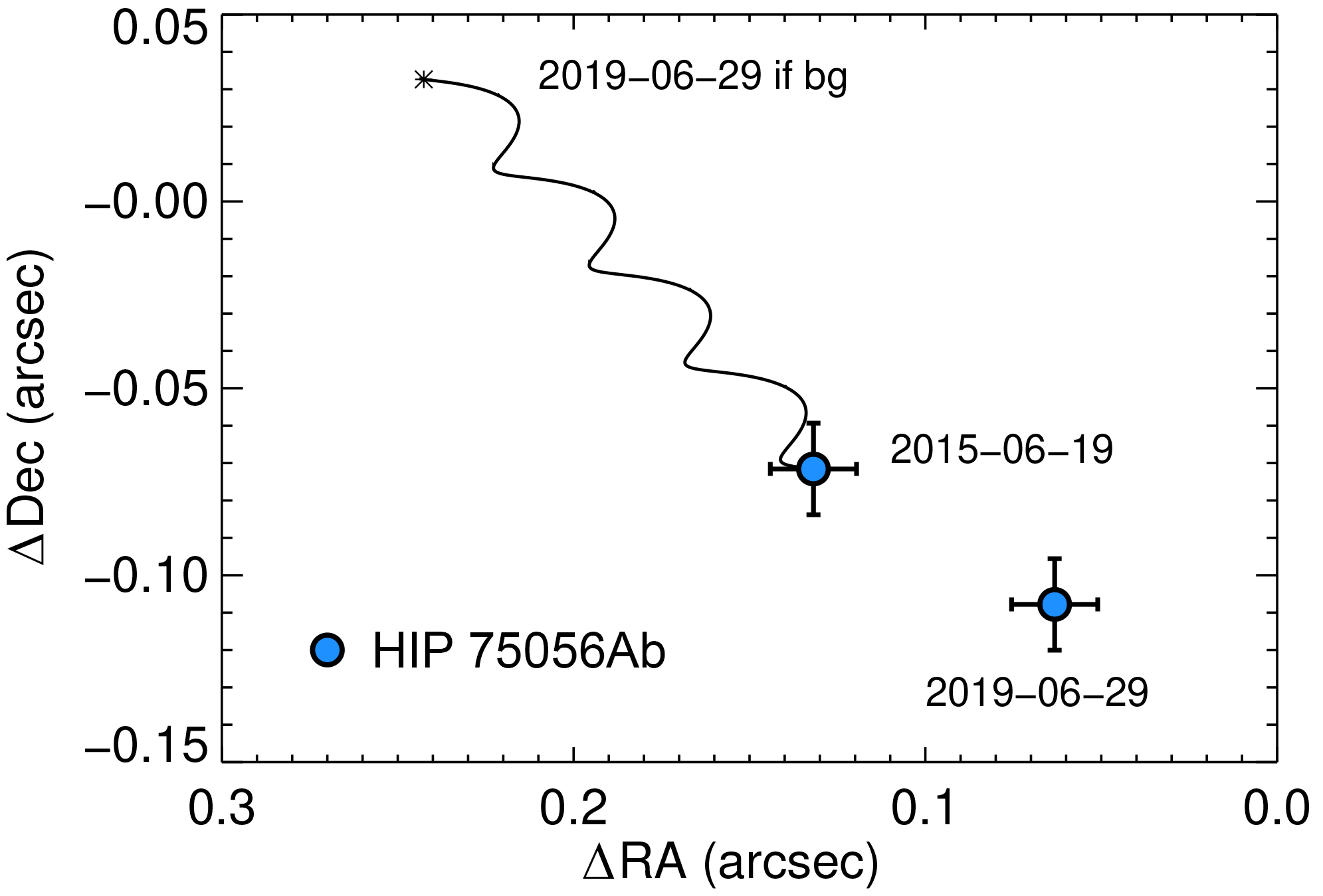}{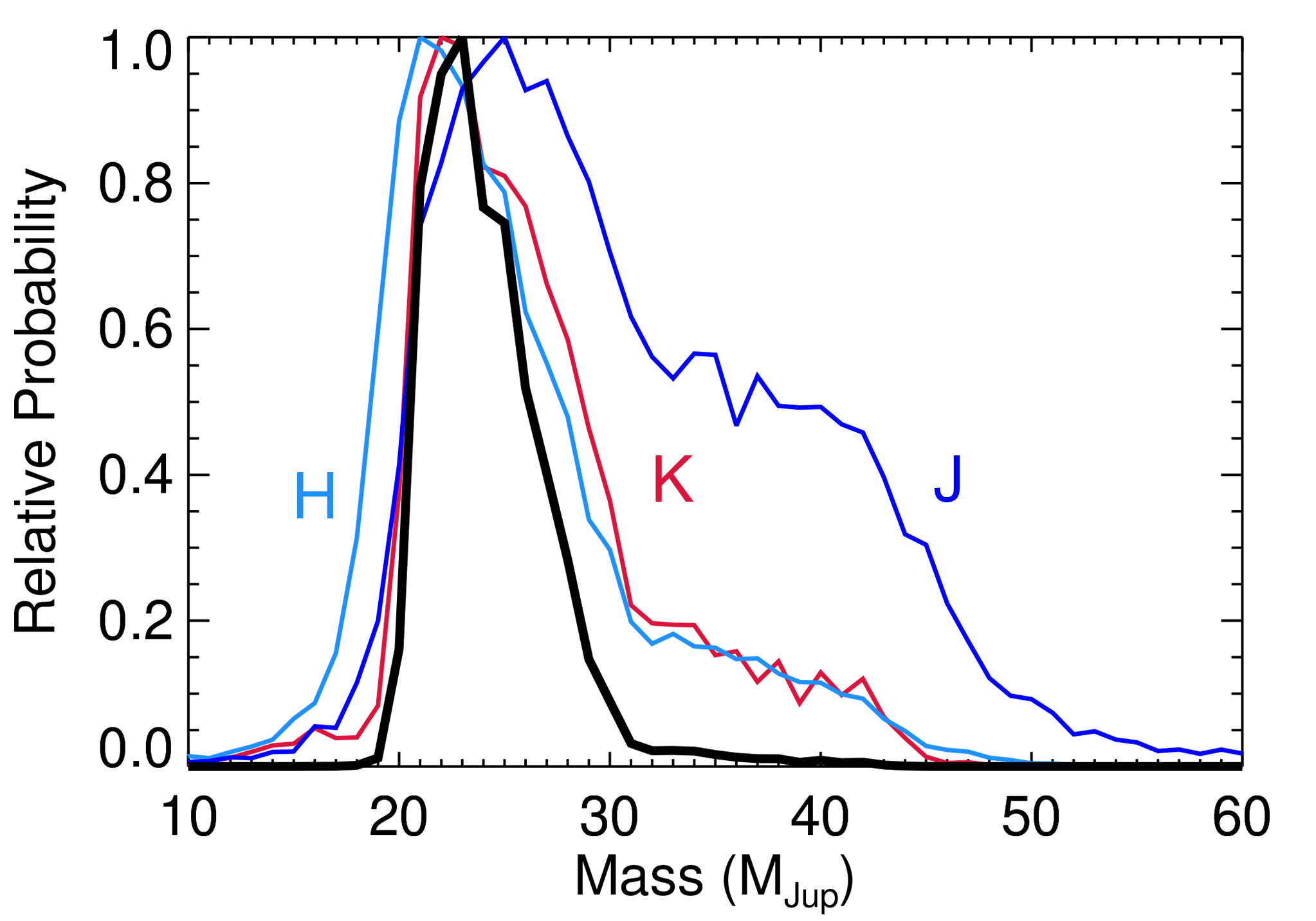}
\caption{Left: Astrometric measurements for HIP 75056Ab. The expected background track is shown in the black curve. HIP 75056Ab moves in the opposite direction of the background, i.e., in the direction of HIP 75056. The remaining motion is consistent with orbital motion of a companion with a semi-major axis of a=30$\pm$15 au. Uncertainties are displayed as 2$\sigma$ (1 pixel) for clarity. Right: Mass estimates for HIP 75056Ab. The combined probability distribution from the three photometric bands is shown in the black curve.}
\end{figure*}

\section{Results}

The images of HIP 75056A are shown in Figure 1. A companion candidate is clearly identified to the SE of HIP 75056A at a projected separation of $\sim$0$\farcs$15 (0$\farcs$125 in 2019). Between 2015 and 2019, the candidate moves around the star to the SW, which is in the same direction as the proper motion of HIP 75056 \citep{GDR2}. A background star, which follows the expected relative motion track to the NE, is also identifiable in the IRDIS K1 images (labeled as ``bg" in the images). Two other background stars with similar relative motions are also present in the full-frame image, as well as HIP 75056B in 2015 (see \S3.4 and Table 1).

\begin{deluxetable}{ccc}
\tablecaption{Properties of HIP 75056}
\tablewidth{8cm}
\tablehead{\colhead{Parameter} & \colhead{Value} & \colhead{Ref.}}
\multicolumn{3}{c}{HIP 75056A}\\
\hline\\
Mass & $\sim$1.92 M$_\odot$ &  1 \\
Age & $\sim$12 Myr & 2\\
Distance & 126 $\pm$2 pc & 3\\
$J$ & 7.38$\pm$0.02 & 4\\
$H$ & 7.35$\pm$0.04 & 4\\
$K$ & 7.30$\pm$0.03 & 4\\
P.M. RA & -22.26$\pm$0.12 mas/yr & 3\\
P.M. Dec & -26.02$\pm$0.10 mas/yr & 3\\

\hline\\
\multicolumn{3}{c}{HIP 75056B}\\
\hline\\
Mass & $\sim$0.3 M$_\odot$ &  1 \\
$q$ & $\sim$0.156  &  1 \\
$\rho(2000)$& 5$\farcs$19 & 1 \\
$\theta(2000)$ & 35$^o$ & 1 \\
$\rho(2015)$& 5$\farcs$159$\pm$0$\farcs$003 & This work \\
$\theta(2015)$ & 33.9$^o \pm0.2^o$ & This work \\
$P$ & 8000$\pm$300 yr & This work \\

\hline\\
\multicolumn{3}{c}{HIP 75056Ab}\\
\hline\\
Parameter & Value & Uncertainty\\
\hline\\
$\Delta J$  & 7.30  & 0.25  \\
$\Delta H$  & 7.15 & 0.25 \\
$\Delta K1$  & 6.80 & 0.10\\
$\Delta K2$  & 6.75 & 0.10\\

M & 25 M$_{Jup}$ & 5 M$_{Jup}$\\
$q$ & 0.012  &  0.002 \\
R  & 2 $R_{Jup}$ & 0.5 $R_{Jup}$ \\
T$_{\rm eff}$  & 2300 K & 300 K \\
SpT & M6$-$L2 & \\
$log$(L/L$_{\odot}$) & -2.83 & 0.07 \\
$\rho$(2015) & 0$\farcs$150 & 0$\farcs$006 \\
$\rho$(2019) & 0$\farcs$125 & 0$\farcs$006 \\
$\theta$(2015) & 118.5$^o$ & 1.3$^o$ \\
$\theta$(2019) & 149.6$^o$ & 1.3$^o$ \\
$a$ & 30 au & 15 au\\
$e$ & 0.5 & 0.2\\
$i$ & 23$^o$ & 11$^o$\\
$P$ & 130 yr & 90 yr
\enddata
\tablecomments{The mass, radius, and temperature of HIP 75056Ab were estimated from a comparison of the photometric measurements to the \cite{Baraffe2003} evolutionary tracks, while its luminosity was estimated based on the conversions in \cite{Golimowski2004, Todorov2010}. References: (1) \citealt{Kouwenhoven2005}, (2) \citealt{Pecaut2016}, (3) \citealt{GDR2}, (4) \citealt{Cutri2003}.}
\end{deluxetable}

\subsection{Astrometry}

The astrometric measurements of HIP 75056Ab display a significant amount of orbital motion ($\sim$10\% of a 45-year orbit: see Table 1 \& Figure 2). Given the large amount of orbital motion, we conservatively estimate the astrometric uncertainties as $\pm$0.5 pixel, or $\pm$0$\farcs$006, and note that these could be overestimated by a factor of $\sim$2 (a complete breakdown of the astrometric error budget for SPHERE can be found in \citealt{Wagner2018}). Over the four-year baseline, we measure a positional shift of $\Delta \rho = 0\farcs025\pm0\farcs008$ and $\Delta \theta = 31.4^o\pm1.8^o$, which is consistent with an orbital period of $\gtrsim$40 years. To place preliminary constraints on the orbital parameters of HIP 75056Ab, we utilized the \texttt{OFTI} method \citep{Blunt2017} in the \texttt{orbitize!} package \citep{Blunt2020}. We generated 10,000 sample orbits consistent with the astrometric data and found average orbital parameters of $a$=30$\pm$15 au, $e$=0.5$\pm$0.2, and $i$=23$^o\pm$11$^o$. The largest semi-major axis would suggest a period of $\sim$ 220 yr. 

With only two data points available, the posterior orbital parameters have significant uncertainties, and are subject to complex degeneracies (e.g., between the inclination, eccentricity, and semi-major axis). For longer period orbits, to observe such a significant change in position angle ($|sim$30$^\circ$) over four years would require the companion to be currently near periastron on an eccentric orbit. The latter scenario is less likely since such a companion would spend the majority of its time at wider separations. With a third epoch observation, these orbital parameters can be significantly improved. With only two epochs available, we caution that this preliminary orbital fit may also be biased by systematics.

\begin{figure*}[htpb]
\epsscale{1.2}
\plotone{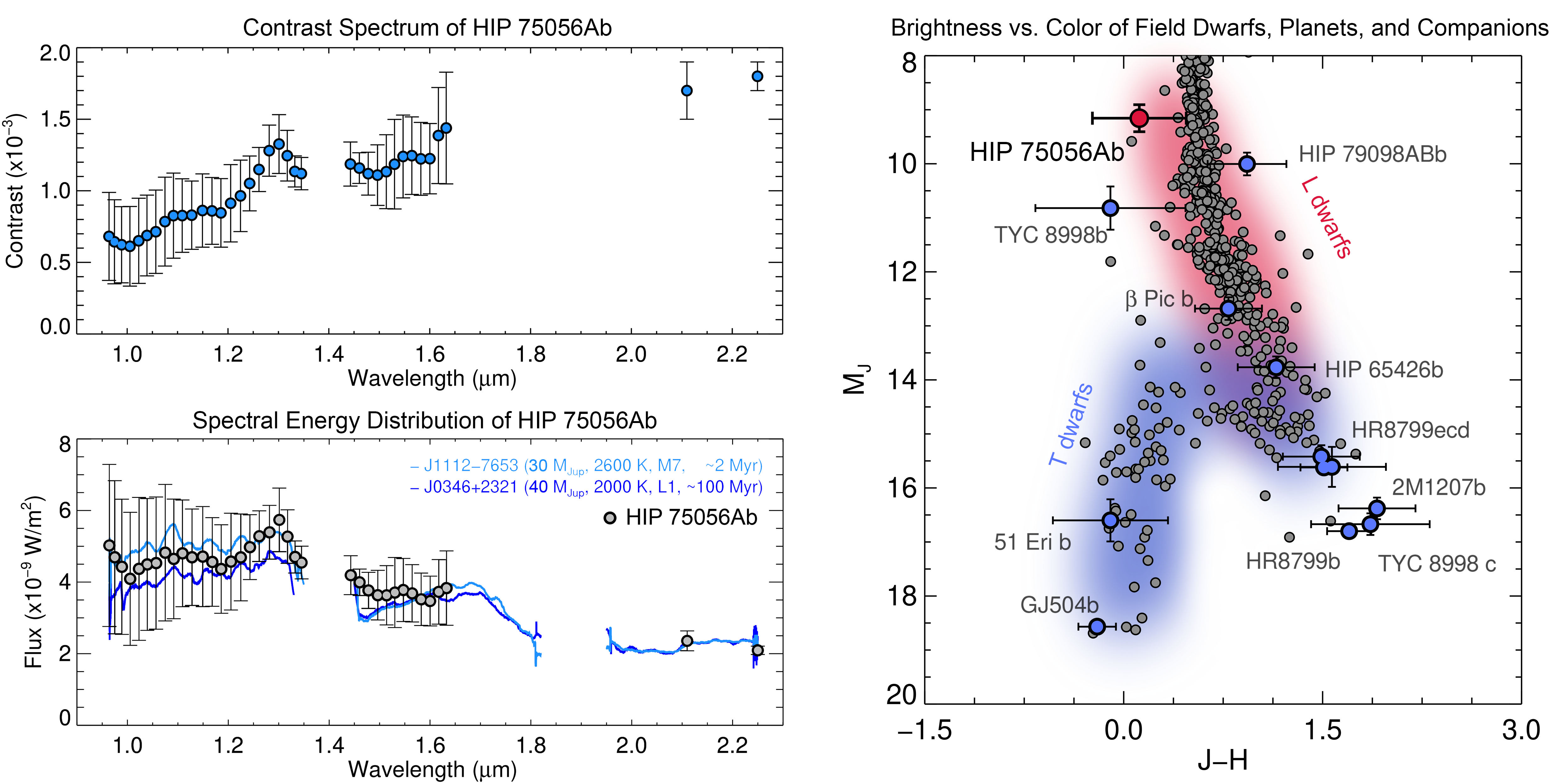}
\caption{Left: Contrast spectrum of HIP 75056Ab with respect to HIP 75056A and spectral energy distribution of HIP 75056Ab. Spectra of young brown dwarfs from \cite{Manjavacas2020} are shown in blue and light blue for comparison. Right: color magnitude diagram of directly imaged planets and field brown dwarfs (gray points). HIP 75056Ab is consistent with a spectral type of M6$-$L2 and a temperature of 2300$\pm$300 K. Notably, HIP 75056Ab is among the youngest and least massive known companions near the M/L transition.}
\end{figure*}

\subsection{Mass Estimates}

We converted the photometric measurements, age, and distance into mass estimates via the evolutionary grids of \cite{Baraffe2003} following the methodology in \cite{Wagner2019}. We converted the $J-$, $H-$, and $K$-band photometry separately into mass estimates, and compute the combined mass probability distribution as the product of the individual distributions. We assumed that the star's $K1$ magnitude is equivalent to its $K$-band magnitude, and also assumed an age of 12 Myr and a 1$\sigma$ age uncertainty of $\pm$5 Myr based on the star's position and the age gradient map in \cite{Pecaut2016}. The results are consistent with a companion mass of $\sim$20$-$30 M$_{Jup}$. Figure 2 illustrates the range of plausible masses and the relative contributions of the photometric bands to the combined probability distribution. The mass range derived from the $J$-band photometry is consistent with somewhat higher masses ($\lesssim$50 M$_{Jup}$), while the $H-$ and $K$-band photometry suggest that the companion's mass is $\lesssim$30 M$_{Jup}$. Since the majority of the mass range is above the deuterium burning limit, the assumption of a high initial planetary entropy does not significantly affect the mass estimates (e.g., \citealt{Mordasini2017, Marleau2019}). We also verified that the atmospheric dust content does not affect the mass estimates by utilizing the model grids of \cite{Chabrier2000}, which produced nearly identical results.

\newpage

\subsection{Spectroscopy}

For the IFS data, we measured the spectrum and uncertainties of HIP 75056Ab using the mean and standard deviation of aperture photometric measurements corrected by the forward modelled spectrum of the star from both ADI-KLIP and ADI+SDI-KLIP for each night (i.e., the mean and standard deviation of four spectra). The synthetic sources were injected at the same separation as HIP 75056Ab, with 8$\times$10$^{-4}$ contrast with respect to HIP 75056A, and at $\Delta$PA = $-$90$^o$, 90$^o$, and 180$^o$ from HIP 75056Ab. For IRDIS, we measured the brightness and position by injecting a negative PSF of the star at the position of the companion. This method provides robust results in the presence of significant self-subtraction and over-subtraction due to ADI. We iterated upon the source's brightness and location, and adopted the parameters that minimized the squared residuals in the final image. The results are shown in Table 1 and Figure 3. We estimated the photometric uncertainties as the standard deviation of the brightness of positive sources injected at the same separation and brightness as HIP 75056Ab and at $\Delta$PA = $-$90$^\circ$, 90$^\circ$, \& 180$^\circ$.

The spectrum of HIP 75056Ab shows a consistent trend of increasing contrast (compared to HIP 75056A) with wavelength. We converted this contrast spectrum into physical flux units by multiplying by a synthetic $T$=7500K, $R=1.9 R_{\odot}$ spectrum \citep{Kurucz1979} scaled to the distance of HIP 75056, which we selected to match the stellar photometry available from 2MASS \citep{Cutri2003}. We visually compared the spectrum to those of young brown dwarfs in \cite{Manjavacas2020}. J1112-7653 and J0346+2321 were chosen among the objects within this library as the two whose spectra appeared most similar to that of HIP 75056Ab. Other spectra within the sample are consistent with the observed $Y-$ to $H$-band spectrum of HIP 75056Ab, but are inconsistent with the full $Y-$ to $K$-band spectrum. These objects have spectral types of M7 and L1, respectively. Since the (normalized) spectra of these objects bracket that of HIP 75056Ab, we estimate a most likely spectral type within the range of M8$-$L0, and conservatively within the range of M6$-$L2. Similarly, we estimate that HIP 75056Ab has an effective temperature of $\sim$2300$\pm$300 K. These are consistent with the evolutionary tracks of \cite{Baraffe2003} and \cite{Allard2012} for a $\sim$20$-$30 $M_{Jup}$ and $\sim$10$-$20 Myr old companion.

\begin{figure*}[htpb]
\epsscale{1.15}
\plotone{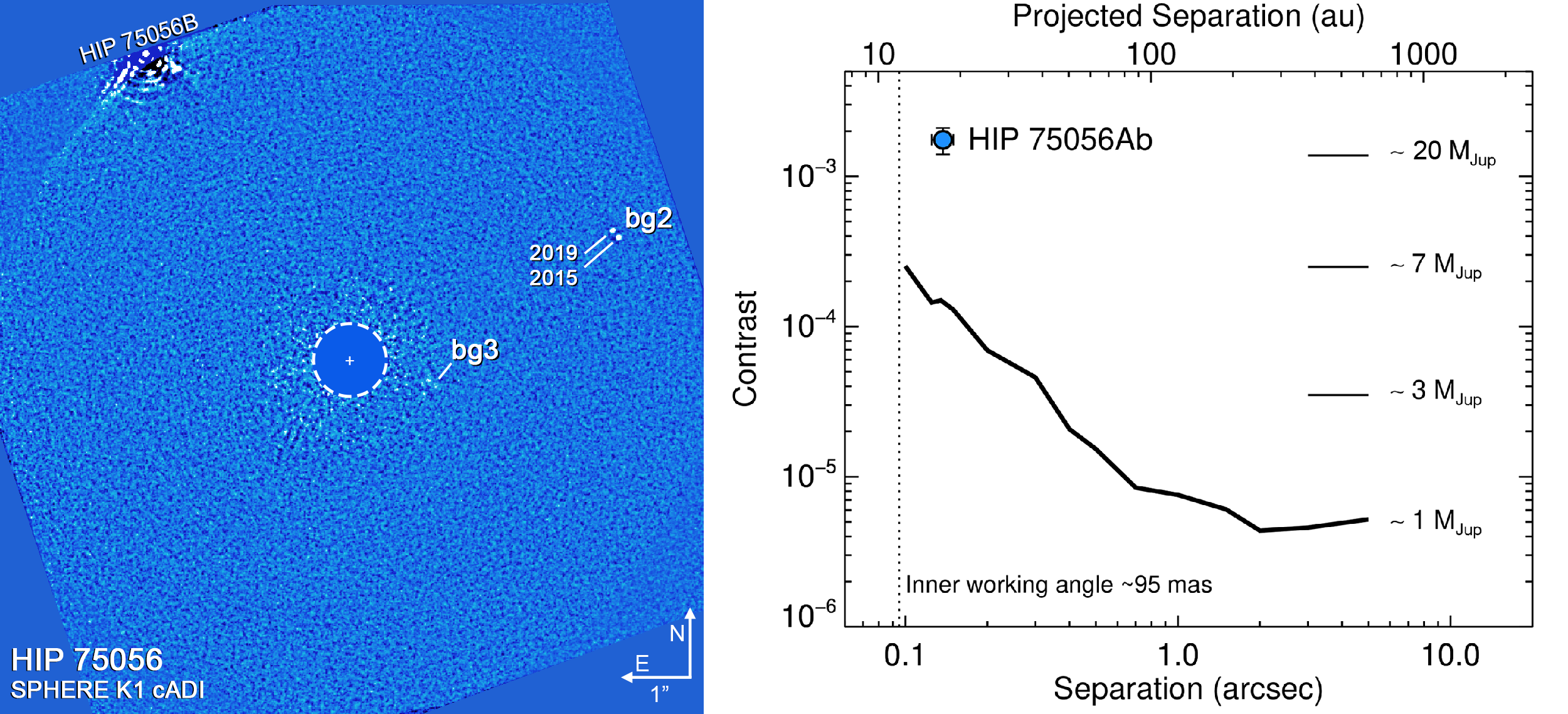}
\caption{Left: full-frame IRDIS $K1$ image processed with cADI. Right: Sensitivity of the 2019 IRDIS $K1$ data processed with KLIP. The average 5$\sigma$ sensitivity estimated from simulated planet injections is shown in the black curve, while the horizontal lines show the corresponding mass estimates from the \cite{Baraffe2003} models.}
\end{figure*}

We also compared the object's brightness and colors to those of field brown dwarfs with known distances \citep{Dupuy2013, Winters2015} and directly imaged planets \citep{Chauvin2005, Lagrange2010, Marois2008, Marois2010, Kuzuhara2013, Macintosh2015, Chauvin2017, Janson2019, Bohn2020, Bohn2020b}. Consistent with the above findings, HIP 75056Ab is positioned in the $J$ vs. $J-H$ color-magnitude diagram near the M/L transition. As one of few known young and low-mass companions near the M/L transition, HIP 75056Ab is potentially useful for comparative atmospheric studies (e.g., \citealt{Sing2016, Madhu2019}).
HIP 75056Ab appears relatively blue among other M/L-transition objects. Notably, TYC 8998b$-$another low-mass companion to a Sco-Cen star$-$is also relatively blue. 

\subsection{Sensitivity Analysis}

We computed the image sensitivity as a function of angular separation from HIP 75056A using simulated point source injection and retrieval tests. We utilized the IRDIS $K1$ image from 2019, which reaches the deepest sensitivity in terms of planetary masses. We measured the signal to noise ratio (SNR) using non-overlapping apertures of $\lambda /D$ diameter, with the starting aperture centered on the position of the injected source, and using the definition of SNR in \cite{Mawet2014}. We excluded a 12x12 pixel region centered on HIP 75056Ab, which would otherwise bias the SNR estimation. We utilized the off-axis PSF of the stars, normalized by the neutral density filter transmission, coronagraph transmission, and difference in exposure time. We converted contrast sensitivities to mass estimates following \S3.2. The results are shown in Figure 4. At the separation of HIP 75056Ab ($\sim$0$\farcs$13), our observations are sensitive to contrasts of $\sim$2$-$3$\times$10$^{-4}$, corresponding to masses of $\sim$7$-$8 M$_{Jup}$. At twice the companion's separation, our observations are sensitive to contrasts of $\sim$3$-$4$\times$10$^{-5}$, or masses of $\sim$2$-$4 M$_{Jup}$. Our observations reach a minimum in contrast sensitivity of $\sim$4$-$5$\times$10$^{-6}$ at separations of $\gtrsim$2", or masses of $\sim$1 M$_{Jup}$. We note that orbits exterior to $\sim$1$\farcs$5 may be unstable due to the effects of HIP 75056B, if the binary is on a nearly face-on and circular orbit \citep{Holman1999}.\\

\section{Discussion}

The discovery of HIP 75056Ab is interesting in two important contexts: 1) providing a template of low surface gravity atmospheres at the M/L transition for spectroscopic studies compared to higher-mass field brown dwarfs and lower-mass giant planets; and 2) establishing the formation mechanism of the companion, which, based on its mass and currently known orbital properties, could be one of the few relatively low-mass  companions formed via disk instability (e.g., \citealt{Boss1997, Kratter2010, Forgan2018}). On the other hand, HIP 75056Ab may represent the high-mass end of the distribution of planets formed via core accretion (e.g., \citealt{Pollack1996, Mordasini2012, Schlaufman2018, Wagner2019}). Establishing the formation mechanism of the companion will also aid in establishing a framework of atmospheric properties for companions formed via different processes.

\subsection{Spectral Analysis}

The spectrum of HIP 75056Ab is similar to that of young brown dwarfs with spectral types of $\sim$M6$-$L2, with the large spread caused by the uncertainties in the IFS data. The CO band-head is present at $\sim$1.3 $\mu$m, as CO is the dominant carbon carrier in M/L-type atmospheres. Compared to field brown dwarfs and directly imaged (young super-Jupiter) exoplanets, HIP 75056Ab appears similar to other objects near the deuterium burning limit, such as TYC 8998b (\citealt{Bohn2020}), a 14$\pm$3 M$_{Jup}$ companion orbiting a sun-like star. Notably, HIP 75056Ab is among the least massive companions discovered to date near the M/L transition. As such, it constitutes a young, hot analogue to colder directly imaged companions of similar masses around older stars, and also a low-mass and low-gravity analogue to older field brown dwarfs of similar temperature.

\subsection{Planet Formation Mechanisms}

The most peculiar aspect of HIP 75056Ab is arguably its formation mechanism. The companion's mass ratio of $q\sim$0.01 is analogous to a $\sim$10 M$_{Jup}$ companion/planet around a Sun-like star. A $\sim$20$-$30 M$_{Jup}$ companion at $\sim$15$-$45 au could represent a rare formation via gravitational instabilities within the protoplanetary disk (see \citealt{Kratter2010, Wagner2019, Tokovinin2020}). Alternatively, HIP 75056Ab might instead be among the most massive companions formed via core accretion (e.g., \citealt{Emsenhuber2020a, Emsenhuber2020b}). Such rare examples often provide the most powerful constraints on formation models (e.g., the four super-Jupiter planets around HR 8799 that are difficult to explain with any mechanism: \citealt{Marois2008, Marois2010}). \cite{Nielsen2019} and \cite{Vigan2020} observed on the order of 100 stars with masses similar to HIP 75056A and found no companions with $q\sim$0.01 and $a\sim$10$-$50 au, suggesting an occurrence rate $\lesssim$1\%. The most similar comparisons to HIP75056Ab are likely the several companions to B/A/F stars in Sco-Cen with $q\sim0.01-0.08$ and $a\sim10-30$ au that were discovered via sparse aperture masking \citep{Hinkley2015}. 



Based on HIP 75056b's mass, it seems most likely that HIP 75056Ab formed via gravitational instability \citep{Wagner2019}; however, a core accretion origin cannot be excluded, and thus it is appropriate to consider another diagnostic. The best indicator is perhaps the companion's orbital eccentricity. \cite{Bowler2020} studied the eccentricity distribution of 27 companions spanning a few Jupiter masses to high-mass brown dwarfs between 5$-$100 au. They found that low mass-ratio companions (i.e. those formed predominantly via core accretion) have typically lower eccentricities than high mass-ratio companions, which have a broad peak at $e\sim$0.6$-$0.9, similar to the wide-binary population. Thus, if HIP 75056Ab is on a high-eccentricity orbit, this would support the hypothesis that it likely formed via gravitational instability (although see \citealt{Emsenhuber2020b}, which shows that high-eccentricity companions might also be formed via core accretion). 

Our two currently available astrometric measurements (spanning four years) are sufficient to place preliminary constraints on the companion's orbit, which suggests $e\sim$ 0.5$\pm$0.2. This is a significant eccentricity compared to other low-mass ratio companions \citep{Bowler2020}, supporting the hypothesis that HIP 75056Ab formed via disk instability (although this does not completely exclude a core accretion origin, see \citealt{Emsenhuber2020b}). In the coming years, continued astrometric monitoring of HIP 75056Ab will be able to verify and better constrain the orbital eccentricity. 

 
Regardless of formation mechanism, given its mass, HIP 75056Ab likely formed early in the system's lifetime–in the Class 0 or I stage when the protoplanetary disk was still massive and embedded in a gaseous envelope \citep{Tychoniec2020}. With a mass accretion rate typical for Class 0/I stars of $\sim$10$^{-5}$ M$_{\odot}$ yr$^{-1}$, massive disks should be unstable at 30 AU \citep{Armitage}. Such early planet formation could lead to an observable signature in the C/O ratio of its atmosphere as compared to field brown dwarfs and giant planets. \cite{vantHoff2020} studied the temperature structures in embedded disks and found that Class 0 disks are warm, with no CO ice frozen out at all and H$_{2}$O ice frozen out only at radii of $\gtrsim$80-100 au. By the Class I stage the disks cool, moving the water snow line further in ($\lesssim$30 au), while allowing CO to freeze out beyond HIP 75056Ab's observed location. If HIP 75056Ab formed in the Class 0/I stage, then it likely did so in the presence of CO in the gas phase and with a decreasing amount of gaseous H$_{2}$O present with time. Therefore, observing a low C/O ratio in its atmosphere may imply an earlier period of runaway gas accretion.
 

\subsection{Images of Companions at Small Separations}

HIP 75056Ab is among several substellar companions that have been directly imaged at very small angular separations (e.g., \citealt{Strampeli2020}). $\beta$ Pictoris b \citep{Lagrange2010, Wang2016}, PDS 70c \citep{Haffert2019}, and HD 206893B \citep{Milli2017} have been imaged interior to 0$\farcs$3. HD 984B \citep{Meshkat2015} and PDS 70b \citep{Keppler2018, Wagner2018b} have been imaged at $\sim$0$\farcs$2. The images of these companions, including HIP 75056Ab as a notable example at 0$\farcs$125, showcase the trend of extreme adaptive optics systems (e.g., \citealt{Macintosh2018, Beuzit2019}) progressing toward imaging companions at smaller angular separations. These separations open interesting possibilities, as an Earth-like planet orbiting in the habitable zone of a Sun-like star at 10 pc would appear at an angular separation of $\sim$0$\farcs$1. 

\section{Summary and Conclusions}

We observed HIP 75056A, an A2V star in the Scorpius-Centaurus OB2 association, with VLT/SPHERE in 2015 and 2019. We detected a companion candidate, HIP 75056Ab, at a small projected separation (0$\farcs$125$-$0$\farcs$15) and a contrast of $\Delta K1=$ 6.8$\pm$0.1 with respect to the primary star. 

We established that HIP 75056Ab is co-moving to the SW along with HIP 75056, and that the object's additional velocity to the SW is consistent with orbital motion for a semi-major axis of 30$\pm$15 au.

We converted HIP 75056Ab's photometric measurements into mass estimates, and obtained consistent results for each photometric band. The combined probability distribution suggests a mass of $\sim$20$-$30 M$_{Jup}$. 

We compared HIP 75056Ab's 0.95$-$2.25 $\mu m$ spectral energy distribution and photometric measurements to young brown dwarfs and directly imaged planets. We found that HIP 75056Ab is consistent with a spectral type of $\sim$M6$-$L2, and a temperature of 2000$-$2600 K. HIP 75056Ab is among the least massive known companions near the M/L transition, making it a useful object for comparative atmospheric studies.

We discussed possible formation mechanisms for HIP 75056Ab, and found that the companion likely formed via gravitational instability, although formation of the companion via core accretion cannot be excluded. Future astrometric measurements of HIP 75056Ab will be able to place better constraints on its orbital parameters$-$in particular its eccentricity, which will help to verify this hypothesis. 

Finally, HIP 75056Ab's detection at 0$\farcs$125 represents a milestone in detecting low-mass companions at separations analogous to the habitable zones of Sun-like stars within 10 pc.

\section{Acknowledgments} The authors acknowledge and thank Kaitlin Kratter and Maxwell Moe for useful conversations regarding the plausible formation scenarios. Support for this work was provided by NASA through the NASA Hubble Fellowship grant HST-HF2-51472.001-A awarded by the Space Telescope Science Institute, which is operated by the Association of Universities for Research in Astronomy, Incorporated, under NASA contract NAS5-26555. The results reported herein benefited from collaborations and/or information exchange within NASA's Nexus for Exoplanet System Science (NExSS) research coordination network sponsored by NASA's Science Mission Directorate.




\end{document}